\documentclass[prb,
preprint,
superscriptaddress, 
secnumarabic, 
amssymb, 
aps,
longbibliography
]{revtex4-2}

\usepackage{bm}
\usepackage{graphicx}
\usepackage{gensymb}
\usepackage{amsmath}
\usepackage{amssymb}
\usepackage{color}
\usepackage{ulem}
\usepackage[driverfallback=dvipdfm]{hyperref}
\hypersetup{%
 setpagesize=false,%
 bookmarks=true,%
 bookmarksdepth=tocdepth,%
 bookmarksnumbered=true,%
 colorlinks=true,%
 allcolors=blue,%
 pdftitle={},%
 pdfsubject={},%
 pdfauthor={},%
 pdfkeywords={}}

\begin{document}

\title{Universal role of combined symmetry for the protection of the Dirac cone in antiferromagnetic topological insulators}%

\author{Asuka Honma}
\affiliation{Department of Physics, Graduate School of Science, Tohoku University, Sendai 980-8578, Japan}

\author{Noriyuki Kabeya}
\affiliation{Department of Physics, Graduate School of Science, Tohoku University, Sendai 980-8578, Japan}

\author{Seigo Souma}
\thanks{Corresponding authors:\\
s.souma@arpes.phys.tohoku.ac.jp\\
t-sato@arpes.phys.tohoku.ac.jp}
\affiliation{Center for Science and Innovation in Spintronics (CSIS), Tohoku University, Sendai 980-8577, Japan}
\affiliation{Advanced Institute for Materials Research (WPI-AIMR), Tohoku University, Sendai 980-8577, Japan}

\author{Yongjian Wang}
\affiliation{Institute of Physics II, University of Cologne, K$\ddot{o}$ln 50937, Germany}

\author{Kunihiko Yamauchi}
\affiliation{Center for Spintronics Research Network (CSRN), Osaka University, Toyonaka, Osaka 560-8531, Japan}

\author{Kosuke Nakayama}
\affiliation{Department of Physics, Graduate School of Science, Tohoku University, Sendai 980-8578, Japan}

\author{Daichi Takane}
\affiliation{Department of Physics, Graduate School of Science, Tohoku University, Sendai 980-8578, Japan}

\author{Kenichi Ozawa}
\affiliation{Institute of Materials Structure Science, High Energy Accelerator Research Organization (KEK), Tsukuba, Ibaraki 305-0801, Japan}

\author{Miho Kitamura}
\affiliation{National Institutes for Quantum Science and Technology (QST), Sendai 980-8579, Japan}

\author{Koji Horiba}
\affiliation{National Institutes for Quantum Science and Technology (QST), Sendai 980-8579, Japan}

\author{Hiroshi Kumigashira}
\affiliation{Institute of Multidisciplinary Research for Advanced Materials (IMRAM), Tohoku University, Sendai 980-8577, Japan}

\author{Tamio Oguchi}
\affiliation{Center for Spintronics Research Network (CSRN), Osaka University, Toyonaka, Osaka 560-8531, Japan}

\author{Takashi Takahashi}
\affiliation{Department of Physics, Graduate School of Science, Tohoku University, Sendai 980-8578, Japan}

\author{Noriaki Kimura}
\affiliation{Department of Physics, Graduate School of Science, Tohoku University, Sendai 980-8578, Japan}

\author{Yoichi Ando}
\affiliation{Institute of Physics II, University of Cologne, K$\ddot{o}$ln 50937, Germany}

\author{Takafumi Sato}
\thanks{Corresponding authors:\\
s.souma@arpes.phys.tohoku.ac.jp\\
t-sato@arpes.phys.tohoku.ac.jp}
\affiliation{Department of Physics, Graduate School of Science, Tohoku University, Sendai 980-8578, Japan}
\affiliation{Center for Science and Innovation in Spintronics (CSIS), Tohoku University, Sendai 980-8577, Japan}
\affiliation{Advanced Institute for Materials Research (WPI-AIMR), Tohoku University, Sendai 980-8577, Japan}
\affiliation{International Center for Synchrotron Radiation Innovation Smart (SRIS), Tohoku University, Sendai 980-8577, Japan}
\affiliation{Mathematical Science Center for Co-creative Society (MathCCS), Tohoku University, Sendai 980-8577, Japan}

\date{\today}

\begin{abstract}
Antiferromagnetic topological insulators (AF TIs) are predicted to exhibit exotic physical properties such as gigantic optical and topological magnetoelectric responses. 
While a key to achieving such phenomena relies on how to break the symmetry protecting the Dirac-cone surface state (SS) and acquire the mass of Dirac fermions, the mechanism has yet to be clarified. 
To address this issue, we carried out micro-focused angle-resolved photoemission spectroscopy for GdBi hosting the type-II AF order, and uncovered the stripe-type 2$\times$1 reconstruction of the Fermi surface associated with the AF band folding. 
Intriguingly, in contrast to NdBi with the type-I AF order displaying the surface-selective Dirac-fermion mass, GdBi shows massless behavior irrespective of AF domains due to the robust topological protection. 
These results strongly suggest a crucial role of the $\mathit{\Theta}$\textit{T}$_\mathrm{D}$ (time-reversal and translational) symmetry to create the Dirac-fermion mass in AF TIs.
\end{abstract}

\maketitle

\section{INTRODUCTION}
TIs are characterized by the boundary state showing linearly dispersive energy bands with a band crossing at the Dirac point\ \cite{FuPRL2007, HsiehNature2008}. 
Symmetries that guarantee such a Dirac-band crossing play a key role in realizing various topological quantum phenomena\ \cite{QiPRB2008, TurnerPRB2012, FuPRL2011}. 
For example, in nonmagnetic TIs, the introduction of ferromagnetic (FM) order breaks the time-reversal symmetry (TRS; $\mathit{\Theta}$) and lifts the band degeneracy at the Dirac point, leading to quantum anomalous Hall effect\ \cite{YuScience2010, ChangScience2013, CheckelskyNatPhys2014}. 
Unlike TRS-broken ferromagnets, many antiferromagnets with a periodic arrangement of magnetic moments have a symmetry analogous to the TRS\ \cite{MongPRB2010}. 
This is called \textit{S} symmetry which is the combination of time-reversal operation $\mathit{\Theta}$ and translational operation \textit{T}$_\mathrm{D}$ (\textit{S} = $\mathit{\Theta}$\textit{T}$_\mathrm{D}$) where the \textbf{D} vector connects antiparallel spins so that the \textit{S} symmetry preserves the structure of antiferromagnets. 
The AF TIs with the \textit{S} symmetry at the surface are characterized by the \textit{Z}$_2$ topological invariant and host the Dirac-cone SS as in the case of time-reversal-invariant TIs\ \cite{MongPRB2010, FangPRB2013}. 
Intriguingly, when the \textit{S} symmetry is broken at the surface with a particular mirror index, an energy gap (Dirac gap) opens at the Dirac point and as a result the Dirac fermions become massive. 
This \textit{S}-symmetry-broken surface leads to the emergence of chiral edge modes\ \cite{MongPRB2010, EssinPRL2009, GuoPRB2011, VarnavaPRB2018} and was theoretically predicted to show unique phenomena such as quantized magnetoelectric effect\ \cite{TurnerPRB2012, MongPRB2010, LiNatPhys2010, MarshPRL2019, ZhangPRL2019}, giant optical responses\ \cite{LiNatPhys2010, SivadasPRL2016, WangnpjCompMat2020, FeiPRB2020}, and axion electrodynamics\ \cite{SekineJApplPhys2021, ArmitageSciPostPhys2019}. 
AF TIs also provide a unique opportunity to manipulate the topological properties via controlling the magnetic domains and N\'{e}el vectors\ \cite{WadleyScience2016, TsaiNature2020, SmejkalNatPhys2018}.

Despite these theoretical predictions, it is still experimentally unclear whether or not the \textit{S} symmetry plays an essential role in characterizing the Dirac-cone SS in AF TIs. 
This is highlighted in the case of MnBi$_2$Te$_4$ (MBT) which was proposed as the first material platform of AF TIs\cite{OtrokovNature2019, LiuNatMat2020, DengScience2020}. MBT is a van-der Waals AF insulator containing a septuple layer in a unit cell with a ferromagnetic Mn layer placed at the center layer wherein adjacent Mn layers are antiferromagnetically coupled. 
At the surface parallel to the septuple layer obtained by the cleaving of the crystal, the \textit{S} symmetry is broken, resulting in the opening of a Dirac gap of ~70 meV\ \cite{OtrokovNature2019}. 
On the other hand, it is hard to access the \textit{S}-symmetry-preserved surface because the cleaving plane is limited to that parallel to the layer. 
This hinders the understanding of the relationship between the \textit{S} symmetry and the Dirac gap.

Recently, rare-earth monopnictide NdBi was proposed to be an AF TI characterized by the topological SS protected by the \textit{S} symmetry\ \cite{HonmaNatCommun2023}. 
NdBi has a cubic rock-salt crystal structure and shows an inverted bulk-band structure that supports the strong TI nature in the paramagnetic (PM) phase. 
Below the N\'{e}el temperature ($T_\mathrm{N} = 24\ \mathrm{K}$), the type-I AF order with the propagation vectors of [100], [010], and [001] directions (equivalent to $<001>$ in a cubic lattice) is stabilized, creating three different types of AF domains on the cleaved (001) surface. 
Angle-resolved photoemission spectroscopy (ARPES) study has identified the AF-domain-dependent Dirac-fermion mass, supporting the protection of Dirac cones by the \textit{S} symmetry\ \cite{HonmaNatCommun2023}. 
On the other hand, an unusual SS, distinct from the Dirac cone, has been observed in NdBi and other type-I AF monopnictides, requiring deeper understanding of the role of symmetries in magnetic topological materials\ \cite{SchrunkNature2022, HonmaPRB2023, KushnirenkoPRB2023}. 
A key to resolve such a complex situation on the interplay between the symmetry and the surface band structure lies in the investigation of topological SS for materials with different AF structures. 

In this study, we deliberately chose GdBi as a target material of ARPES investigation, because it exhibits the type-II AF order [Fig.\ \hyperref[FIG1]{1(a)}] with the propagation vector along the $<111>$ direction\ \cite{GdBi111ref1}, distinct from the type-I order of NdBi with the $<001>$ direction. We have found that the Dirac-cone SS remains gapless even in the AF phase, in contrast to the case of NdBi which shows the surface-selective Dirac-fermion mass. We discuss the origin of such a difference in terms of the \textit{S} symmetry.

\section{\label{sec:level1}EXPERIMENTS}
GdBi single crystals were grown by the Bridgman method in a high-vacuum furnace equipped with a tungsten heater. 
High-purity starting materials of Gd (3N) and Bi (6N) with a ratio of 1:1 were sealed in a tungsten crucible using an electron-beam welder. 
The crucible was heated above the melting point of sealed materials (2,100 $^\circ \mathrm{C}$) and then slowly pulled down from the heater. 
After that, the crucible was slowly cooled down to room temperature. Obtained crystals were characterized by x-ray diffraction and magnetization measurements.

Vacuum-ultraviolet (VUV)-ARPES measurements were performed at BL-28A in Photon Factory (PF) with circularly polarized 60–200 eV photons using a micro beam spot of 12 $\times$ 10 $\mathrm{\mu m}^2$\ \cite{KitamuraRSI2022}. 
Soft-x-ray (SX) ARPES measurements were performed at BL-2A (Musashi) in PF with linearly polarized light (horizontal polarization) of 536 eV. 
The energy resolution for VUV and SX-ARPES measurements was set to be 10–30 meV and 150 meV, respectively. 
Samples were cleaved \textit{in situ} in an ultrahigh vacuum of $\sim 1 \times 10^{-10}$ Torr along the (001) crystal plane. 
The Fermi level (\textit{E}$_\mathrm{F}$) of the samples was referenced to that of a gold film evaporated onto the sample holder.
We calibrated the temperature of sample with an accuracy of $\pm$ 1 K.

First-principles band-structure calculations were carried out by using the projector augmented wave method implemented in the Vienna Abinitio Simulation Package (VASP) code\ \cite{KressePRB1996}. 
To calculate the band structure in the PM phase, the modified Becke–Johnson (mBJ) potential\ \cite{BeckeJChemPhys2006}, which is known to properly reproduce the band gap in rare-earth monopnictides\ \cite{LiPRB2018}, was used for the exchange-correlation functional. 
The total energy was calculated self-consistently with the tetrahedron sampling of $8\times8\times1$ \textit{k}-point mesh taking into account spin-orbit coupling. 
The SS was obtained with the surface Green’s function method implemented in WannierTools code\ \cite{WuCompPhysCommun2018} after the maximally localized Wannier functions for Bi-\textit{s}, Bi-\textit{p}, and Gd-\textit{d} orbital states were obtained by using Wannier90 code\ \cite{MostofiCompPhysCommun2008}.

\section{RESULTS AND DISCUSSION}
First, we present a characterization of a GdBi single crystal grown by the Bridgman method. 
Temperature dependence of magnetic susceptibility $\chi (M/H)$ in Fig.\ \hyperref[FIG1]{1(b)} signifies a peak at 27.5 K associated with the AF transition, consistent with the previous report\ \cite{LiJPhysCondMat1997}. 
The $1/ \chi$ shows a linear temperature dependence above $T_\mathrm{N}$ of 27.5 K (inset), suggesting that contributions from core diamagnetism, Pauli paramagnetism, and van Vleck paramagnetism are small. 
As shown in Fig.\ \hyperref[FIG1]{1(c)}, the core-level photoemission spectrum in a wide binding energy ($E_\mathrm{B}$) region is composed only of the Bi- and Gd-derived peaks, and the Laue diffraction pattern shows sharp spots with four-fold symmetry of the (001) cleaving plane. 
These suggest the high-quality nature of our bulk crystal.

To clarify the existence of topological Dirac-cone SS in the PM phase, we carried out surface-sensitive ARPES measurements with VUV photons in a wide valence-band region along the $\bar{\mathrm{\Gamma}}\bar{\mathrm{M}}$ cut of surface Brillouin zone (BZ), as shown in Fig.\ \hyperref[FIG1]{1(d)}. 
The existence of topological SS in the PM phase is a necessary condition for GdBi to become an AF TI below $T_\mathrm{N}$. Due to the spectral broadening along the wave vector perpendicular to the surface associated with the short escape depth of photoelectrons, the ARPES signal for the bulk states reflects the electronic structure integrated over a large $k_z$ area\ \cite{HonmaNatCommun2023, OinumaPRB2019}. 
In the experiment, one can recognize two hole bands centered at the $\bar{\mathrm{\Gamma}}$ point, h1 and h2, showing stronger and weaker intensities, respectively, both of which are ascribed to the bulk Bi 6\textit{p} bands at the $k_z = 0$ plane. 
We also find a shallow and weak electron band e1 which is attributed to the Gd 5\textit{d} band at $k_z = 0$. 
DFT calculations along the $\mathrm{\Gamma}$X cut of bulk BZ (corresponding to the $\bar{\mathrm{\Gamma}}\bar{\mathrm{M}}$ cut in the surface BZ) shown by black curves in Fig.\ \hyperref[FIG1]{1(e)} reproduce well the experimental h1, h2 and e1 bands. 
Importantly, according to the DFT calculations, the h2 and e1 bands show an inverted band structure at the $\bar{\mathrm{M}}$ point producing a small spin-orbit gap at their intersection slightly away from the $\bar{\mathrm{M}}$ point (black arrows; see Appendix\ \hyperref[apxA]{A} for details of the experimental and calculated band dispersions). 
In fact, our parity analysis on the calculated band structure at eight time-reversal-invariant momenta (TRIMs) shows topological invariants of $(\nu_0; \nu_1, \nu_2, \nu_3) = (1; 0, 0, 0)$, supporting that GdBi is a strong $Z_2$ TI with a negative band gap in the PM phase, as in other \textit{R}Bi compounds such as NdBi\ \cite{HonmaNatCommun2023}. 
Associated with the bulk-band inversion at the bulk X point, single (called D1) and double (called D2 and D3) Dirac-cone SSs appear at the $\bar{\mathrm{\Gamma}}$ and $\bar{\mathrm{M}}$ points, respectively, corresponding to the number of band inversions projected onto the surface BZ. 
Such Dirac-cone SSs are identified in both the experiment [Fig.\ \hyperref[FIG1]{1(d)}] and the theoretically-calculated surface spectral weight obtained with the Green’s function method [Fig.\ \hyperref[FIG1]{1(e)}]. 
It is noted that the energy separation between the Dirac points of the D2 and D3 SSs at the $\bar{\mathrm{M}}$ point is very close to each other, and the energy dispersion of the D3 SS is less clear due to the stronger interaction with the bulk states.

To examine the influence of type-II AF order on the fermiology, we carried out Fermi-surface mapping across $T_\mathrm{N}$. 
As shown in Fig.\ \hyperref[FIG2]{2(a)}, the ARPES intensity in the PM phase is dominated by the warped square-like outer hole pocket at the $\bar{\mathrm{\Gamma}}$ point (h2) and the cross-shaped electron pockets (e1 and e2) at the $\bar{\mathrm{M}}$ point, consistent with the semimetallic bulk-band structure shown in Fig.\ \hyperref[FIG2]{2(b)} under the influence of strong $k_z$ broadening (note that the h1 and D1 pockets are inside the h2 pocket and not clearly visible). 
We have examined homogeneity of the observed electronic structure by scanning the whole area of the cleaved surface with the micro beam spot. 
As a result, we found that the intensity pattern is homogeneously distributed over the entire surface in the PM phase, whereas the intensity in the AF phase exhibits two different types of patterns depending on the $(x, y)$ position of the surface. 
We categorize these as domains A and B, respectively, and their distribution is visualized by green and orange colors in Fig.\ \hyperref[FIG2]{2(c)}. 
As shown in Fig.\ \hyperref[FIG2]{2(d)}, the intensity distribution for domain A is apparently different from that of the PM phase [Fig.\ \hyperref[FIG2]{2(a)}]. 
Besides the Fermi surface seen in the PM phase, we found a replica of the $\bar{\mathrm{\Gamma}}$-centered square pocket with a reduced intensity midway between the first and second $\bar{\mathrm{\Gamma}}$ points; specifically, at $(k_x, k_y) = (1.0, -1.0)$ and (-1.0, 1.0) in a unit of $\pi/a$. 
This results in a 1D array of square pockets running along the lower-right/upper-left direction [see also schematics in Fig.\ \hyperref[FIG2]{2(e)}]. 
Such replicas are also observed for the $\bar{\mathrm{M}}$ centered cross-shaped pockets at $(k_x, k_y) = (1.0, 1.0)$ and (-1.0, -1.0), resulting in the stripe-type reconstruction of the Fermi surface. 
In domain B [Figs.\ \hyperref[FIG2]{2(f)} and \ \hyperref[FIG2]{2(g)}], all these replica pockets are resolved, but the direction of 1D array is different; it runs along the lower-left/upper-right direction, rotated by $90^\circ$ from that of domain A. 
This rotation can be better seen by directly looking at the experimental band dispersion, as detailed in Appendix \hyperref[apxB]{B} and Fig. \hyperref[FIG6]{6}.
All these features are well explained in terms of the AF-induced band folding with two types of surface-projected AF ordering vectors depicted by $\mathrm{\mathbf{G}_{AF}}$ (green and orange arrows) in Figs.\ \hyperref[FIG2]{2(d)–2(g)}. 
When we take into account the cubic symmetry of bulk crystal and the type-II AF order, there exist four types of AF domains in the bulk characterized by four AF-ordering vectors, $[111]$, $[\bar{1}11]$,$ [1\bar{1}1]$, and $[\bar{1}\bar{1}1]$ [orange and green arrows in the bulk BZ in Fig.\ \hyperref[FIG2]{2(b)}]. 
When these vectors are projected onto the (001) surface [surface BZ with purple shade in Fig.\ \hyperref[FIG2]{2(b)}], two of them, e.g. $[111]$ and $[\bar{1}\bar{1}1]$ exhibit the same band folding, leading to the distinction of two types of AF domains by ARPES (note that there exist two possible magnetic structures (type A and type B) for the type-II AF order, and the present result supports type A, as detailed in Appendix \hyperref[apxC]{C}).
Despite such strong domain-dependent Fermi-surface modulation, we found that the Dirac-cone band (D1) at the $\bar{\mathrm{\Gamma}}$ point seen in the PM phase remains unchanged even in the AF phase. 
As shown in Figs.\ \hyperref[FIG2]{2(h)} and\ \hyperref[FIG2]{2(i)}, one can commonly recognize the gapless X-shaped D1 band for both domains A and B.
It is noted that we adopted the energy resolution (energy width resolution) of 15 meV which corresponds to the energy position resolution of $\sim$ 2 meV. Thus, the upper limit of the energy gap for the ``gapless'' Dirac cone would be $\sim$ 2 meV.

We have confirmed that the D1 band in the AF phase is indeed of surface origin by observing that the band dispersion is robust against $h\nu$ variation despite a change in the intensity relative to the bulk hole bands h1 and h2 [Figs.\ \hyperref[FIG3]{3(a)}–\hyperref[FIG3]{3(h)}]. 
We also found that the D2 band shows a negligible change across $T_\mathrm{N}$ and preserves the gapless character below $T_\mathrm{N}$ [Fig.\ \hyperref[FIG3]{3(d)}], as in the case of the D1 band [note that it was difficult to trace the whole dispersion of the D3 band in the experiment likely due to the stronger interaction with the bulk states, as inferred from the DFT calculation in Fig.\ \hyperref[FIG1]{1(e)}]. 
Such AF-transition-insensitive nature of the D1 band is distinctly different from the behavior of isostructural NdBi\ \cite{HonmaNatCommun2023} with the type-I AF order showing an energy gap of $\sim125\ \mathrm{meV}$ in the AF phase on the top surface [see also Fig.\ \hyperref[FIG4]{4(c)}], as highlighted by a side-by-side comparison of the experimental D1-band dispersion between GdBi and NdBi in Fig.\ \hyperref[FIG3]{3(e)}. 
This is also evident from the EDC at the $\bar{\mathrm{\Gamma}}$ point [Fig.\ \hyperref[FIG3]{3(f)}], which shows a single peak in GdBi while a two-peaked structure is seen in NdBi. 
Intriguingly, a similar trend was also identified for the D2 band; the D2 band in NdBi shows an energy gap [blue circles in Fig.\ \hyperref[FIG3]{3(g)}] accompanied by the double-peaked feature in the EDC [blue curve in Fig.\ \hyperref[FIG3]{3(h)}], while in GdBi such an energy gap is not clearly seen [red markers in Fig.\ \hyperref[FIG3]{3(g)}] and the EDC shows a single peak [red curve in Fig.\ \hyperref[FIG3]{3(h)}]. 

It is emphasized here that, in FM TIs, the out-of-plane component of magnetic moment at the surface is essential to create a finite Dirac-fermion mass associated with the TRS breaking. 
On the other hand, GdBi with the type-II AF structure always shows a massless Dirac cone at the surface with a finite out-of-plane magnetic component [Fig.\ \hyperref[FIG1]{1(a)}]. 
This strongly suggests that the TRS alone cannot account for the Dirac-cone protection in GdBi, requiring consideration of other symmetries unique to the AF structure.

Now we discuss a critical difference in the nature of the surface Dirac-fermion mass between GdBi and NdBi. 
In the PM phase, both compounds commonly show bulk-band inversion and gapless Dirac-cone SS associated with the $Z_2$ topology on the (001) cleaved surface [Fig.\ \hyperref[FIG4]{4(a)}], and their topological properties are essentially the same with each other. 
On the other hand, in the AF phase, GdBi and NdBi show different (type-II and type-I, respectively) AF order [Figs.\ \hyperref[FIG4]{4(b)} and\ \hyperref[FIG4]{4(c)}]. 
In NdBi, the type-I AF order leads to three types of AF domains on the (001) surface with the magnetic propagation vectors running along the [100], [010], and [001] axes, respectively. 
In a rock-salt crystal, equivalently, one can consider the single AF domain with the fixed ordering vector and three cleaving planes (100), (010), and (001), as shown in Fig.\ \hyperref[FIG4]{4(c)}. 
Since the (100) and (010) surfaces (side surfaces) have the \textbf{D} vector (the translation that inverts the spin direction) lying on the surface, the combined symmetry \textit{S} is preserved, and the Dirac cone keeps the massless character. 
On the other hand, the \textit{S} symmetry is broken on the (001) surface (top surface) and the Dirac cone becomes massive because the \textbf{D} vector cannot be defined parallel to the surface. 
This leads to the weak-TI-like behavior on the Dirac-cone SS in NdBi. 
In GdBi, there exist two types of AF domains in the (001) surface with propagation vectors of [110] and $[1\bar{1}0]$ (see also Fig.\ \hyperref[FIG2]{2}). 
In this case, one can define the \textbf{D} vectors parallel to the surface for both the top and side surfaces [Fig.\ \hyperref[FIG4]{4(b)}], leading to the emergence of gapless Dirac cones and strong-TI-like characteristics. 
This consideration led us to conclude that the \textit{S} symmetry is crucial for determining the massive \textit{vs}.\ massless character of Dirac fermions in AF TIs independently of the AF structures. 
This conclusion is reasonable because only the \textit{S} symmetry is a common symmetry between surface and bulk among various crystal symmetries (e.g. \textit{I}, $C_{3, [111]}$, $\{m_{\bar{1}10}|1,1,1\}$) in the AF phase of GdBi (magnetic space group; $R_\mathrm{I}\bar{3}c$, No. 167.108\ \cite{Perez-MatoARevMatRes2015, StokesISOMAGsite}).

Here we briefly comment on another possibility of the gap opening. A reflective magnetic circular dichroism (RMCD) study of MnBi$_2$Te$_4$  has revealed a nonzero residual magnetism even at zero field \cite{YangPRX2021} and an ARPES measurement observed a gap in the trivial SS \cite{DuySciAdv2024}. These experimental results can be interpreted in terms of surface ferromagnetism. Since RMCD data are not available for NdBi or GdBi at the moment, one cannot completely exclude the possibility that the gap in NdBi is associated with the surface ferromagnetism, although our ARPES results of NdBi and GdBi are more compatible with the \textit{S}-symmetry protection. If surface ferromagnetism is the case, the observed very large gap of  $\sim125\ \mathrm{meV}$ would lead to a large RMCD signal, although this awaits further experimental verification.

The present results confirm the theoretically proposed condition $\mathrm{\mathbf{N}}\cdot\mathrm{\mathbf{D}} = 0$ ($\mathrm{\mathbf{N}}$: a vector normal to the surface) to keep the gapless Dirac-cone SS in AF TIs with the \textit{S} symmetry\ \cite{MongPRB2010, FangPRB2013}. 
Here, the $\mathrm{\mathbf{D}}$ vector hosting the gapless Dirac-cone SS is generally written as, 
\begin{eqnarray}
\mathrm{\mathbf{D}}=\sum^{3}_{i=1}(n_i+\frac{x_i}{2})\mathrm{\mathbf{a}}'_i
\end{eqnarray}
where $\mathrm{\mathbf{a}}'_i$ [shown by open black arrows in Figs.\ \hyperref[FIG4]{4(a)}–\hyperref[FIG4]{4(c)}] and $n_i$ are the unit vector in the AF phase and an arbitrary integer, respectively, and $x_i$ is 0 or 1 (at least one of $x_i$’s must be 1)\ \cite{FangPRB2013}. 
The gapless surfaces of NdBi and GdBi in the AF phase satisfy the above condition, because the shortest \textbf{D} vector for NdBi can be written with the set of $(n_1, n_2, n_3) = (0, 0, 0)$ and $(x_1, x_2, x_3) = (1, 1, 1)$, whereas in GdBi the $(n_1, n_2, n_3)$ and $(x_1, x_2, x_3)$ are (-1, 0, 0) and (1, 1, 1), respectively [orange arrows in Figs.\ \hyperref[FIG4]{4(b)} and\ \hyperref[FIG4]{4(c)}]. 
Thus, in AF TIs, the deliberate choice of a surface with a suitable mirror index can lead to the \textit{S}-symmetry-broken massive Dirac-cone SS which is a prerequisite to realizing $Z_2$-originated optical and electromagnetic responses\ \cite{FangPRB2013, LiNatPhys2010}. 
In this respect, the AF TIs with \textit{S} symmetry are advantageous to realize the exotic quantum phenomena, compared to the FM case which often requires the complex hetero-junction of TI and ferromagnet. 
Such surface engineering utilizing the \textit{S}-symmetry-characterized Dirac-cone SS is the next step to realizing exotic phenomena unique to AF TIs.

\section{CONCLUSION}
We have performed AF-domain-selective micro-ARPES measurements on type-II antiferromagnet GdBi, and uncovered (i) stripe-type reconstruction of Fermi surface associated with the AF band folding and (ii) robust gapless Dirac-cone SS in the AF phase in contrast to surface-selective Dirac gap in isostructural type-I antiferromagnet NdBi. Our result strongly suggests the universal role of \textit{S} symmetry for the protection of the Dirac cone in AF TIs, and paves a pathway toward realizing exotic quantum phenomena in AF TIs.

Note added: after the completion of this work, we became aware of a similar ARPES work by Kushnirenko \textit{et al.} \cite{refKaminski} that commonly observed the reconstruction of Fermi surface associated with the AF order. There are some differences between the two studies on the discussion of the topological aspects of the Dirac-cone SS and the identification of different magnetic domains.

\begin{acknowledgments}
We thank T. Kato, T. Osumi, and Y. Kondo for their assistance in the ARPES experiments. 
This work was supported by JST-CREST (No. JPMJCR18T1), Grant-in-Aid for Scientific Research (JSPS KAKENHI Grant Numbers JP21H04435 and JP24K00564), Grant-in-Aid for JSPS Research Fellow (No: JP23KJ0210), KEK-PF (Proposal number: 2021S2-001, 2024S2-001, and 2022G652), and UVSOR.  
The work in Cologne was funded by the Deutsche Forschungsgemeinschaft (DFG, German Research Foundation) - Project number 277146847 - CRC 1238 (Subproject A04). A.H. thanks GP-Spin and JSPS, and D.T. thanks JSPS and Tohoku University Division for Interdisciplinary Advanced Research and Education.
\end{acknowledgments}

\appendix
\section{BULK BAND INVERSION IN GdBi}
\label{apxA}Bulk-sensitive SX ARPES is advantageous in visualizing the bulk band dispersion of GdBi. Figure\ \hyperref[FIG5]{5(a)} shows raw energy distribution curves (EDCs) along the $\mathrm{\Gamma X}$ cut of bulk BZ measured with $h\nu = 536\ \mathrm{eV}$. The plot signifies a weak Gd 5\textit{d} electron band e1 which crosses $E_\mathrm{F}$ midway between the $\mathrm{\Gamma}$ and X points, together with the inner (h1) and outer (h2) Bi-6\textit{p} hole bands around the $\mathrm{\Gamma}$ point (blue dots). The e1 and h2 bands appear to show a large hybridization gap, suggestive of the inverted band structure as in the case of other \textit{R}Bi compounds\ \cite{OinumaPRB2019, LiPRB2018, HonmaNatCommun2023}. This behavior is also confirmed by corresponding ARPES intensity and its second derivative plots in Figs. \hyperref[FIG5]{5(b)} and \hyperref[FIG5]{5(c)}, respectively.  As shown in Fig.\ \hyperref[FIG5]{5(d)}, the ARPES-derived band dispersions are well reproduced by our band-structure calculations including the spin-orbit coupling which signify the bulk band inversion with a small spin-orbit gap (see black arrows). These results, together with the observation of Dirac-cone SS in the PM phase [Fig.\ \hyperref[FIG1]{1(d)}], support the TI nature of GdBi in the PM phase.

\section{DISTINCTION OF TWO MAGNETIC DOMAINS}
\label{apxB} We show in Figs.\ \hyperref[FIG6]{6(a)} and \hyperref[FIG6]{6(b)} Fermi-surface intensity mapping that signifies weak signature of folded replica Fermi surfaces (same as Figs.\ \hyperref[FIG2]{2(d)} and \hyperref[FIG2]{2(f)}, but without guidelines for the Fermi surface). To show that the AF ordering vector ($\mathrm{\mathbf{G}_{AF}}$) of domain B is perpendicular to that of domain A, we inspected the band dispersion along a $k$ cut passing through the folded bands, as shown by a red dashed line in Figs.\ \hyperref[FIG6]{6(a)} and \hyperref[FIG6]{6(b)}. A side-by-side comparison of Figs.\ \hyperref[FIG6]{6(c)} and \hyperref[FIG6]{6(d)} suggests that the intensity distribution of the folded bands is reversed with respect to $k_x$ = 0, supporting the picture shown in Figs.\ \hyperref[FIG2]{2(e)} and \hyperref[FIG2]{2(f)} where the 1D array of the Fermi surfaces is rotated by 90$^{\circ}$ from each other.

\section{IDENTIFICATION OF MAGNETIC STRUCTURES}
\label{apxC} 
\begin{figure}
\begin{center}
\includegraphics[width=5.5in]{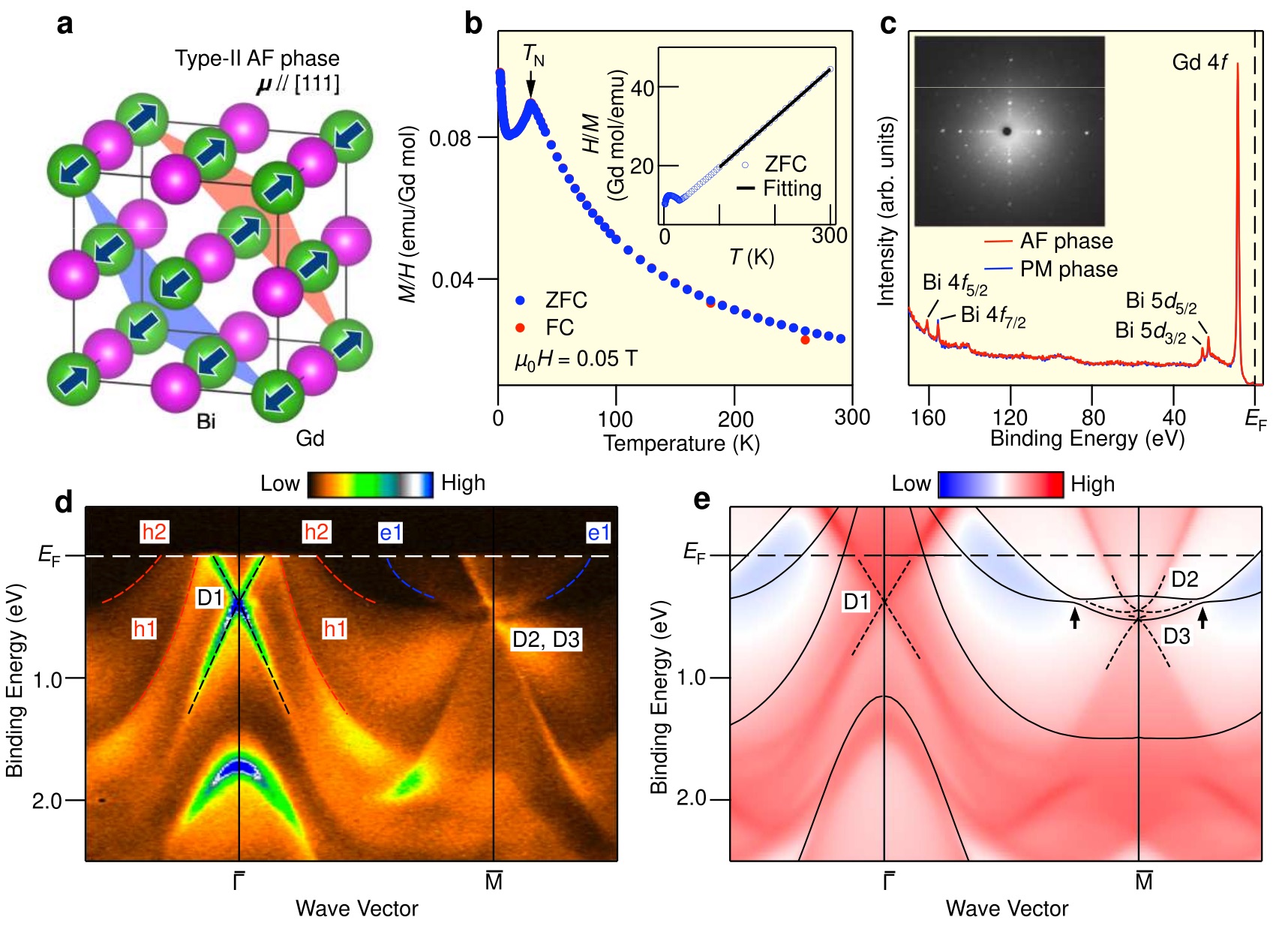}
\caption{\label{FIG1}(color online). (a) Crystal structure and spin configuration of GdBi in the type-II AF phase. FM layers are antiferromagnetically stacked along the $<111>$ direction. (b) Temperature dependence of the $M/H$ curve obtained under the magnetic field along the [001] direction. Blue and red circles represent data obtained with the zero-field cooling (ZFC) and field-cooling (FC) conditions, respectively. Black arrow corresponds to the N\'{e}el temperature ($T_\mathrm{N} = 27.5\ \mathrm{K}$). Inset shows the $H/M$ plot for the ZFC data. Black solid line shows a linear fit based on the Currie-Weiss’s law. The absolute value of effective magnetic moment ($\sim$ 8 $\mu_\mathrm{B}$) estimated from the slope of the curve in the inset is close to the magnetic moment for Gd$^{3+}$ (7.94 $\mu_\mathrm{B}$). (c) Photoemission spectra in a wide $E_\mathrm{B}$ region measured with $h\nu = 200\ \mathrm{eV}$ at $T = 7\ \mathrm{K}$ (AF phase; red curve) and 30 K (PM phase; blue curve). Inset shows the Laue diffraction pattern. (d) Plot of ARPES intensity as a function of $E_\mathrm{B}$ and wave vector along the $\bar{\mathrm{\Gamma}}\bar{\mathrm{M}}$ cut measured in the PM phase ($T = 40\ \mathrm{K}$) with $h\nu = 90\ \mathrm{eV}$. Red, blue, and black dashed curves are a guide for the eyes to trace the Bi 6\textit{p} (h1 and h2), Gd 5\textit{d} (e1), and surface Dirac-cone (D1) bands, respectively. (e) Calculated surface spectral weight along the $\bar{\mathrm{\Gamma}}\bar{\mathrm{M}}$ cut projected onto the (001) plane, obtained with the Green’s function method for a semi-infinite slab of GdBi in the non-magnetic phase. Black dashed curves trace the band dispersion of the D1, D2 and D3 SSs. Black solid curves are calculated bulk band dispersion at $k_z = 0$ ($\mathrm{\Gamma X}$ cut in bulk BZ).}
\end{center}
\end{figure}

\begin{figure}
\begin{center}
\includegraphics[width=4in]{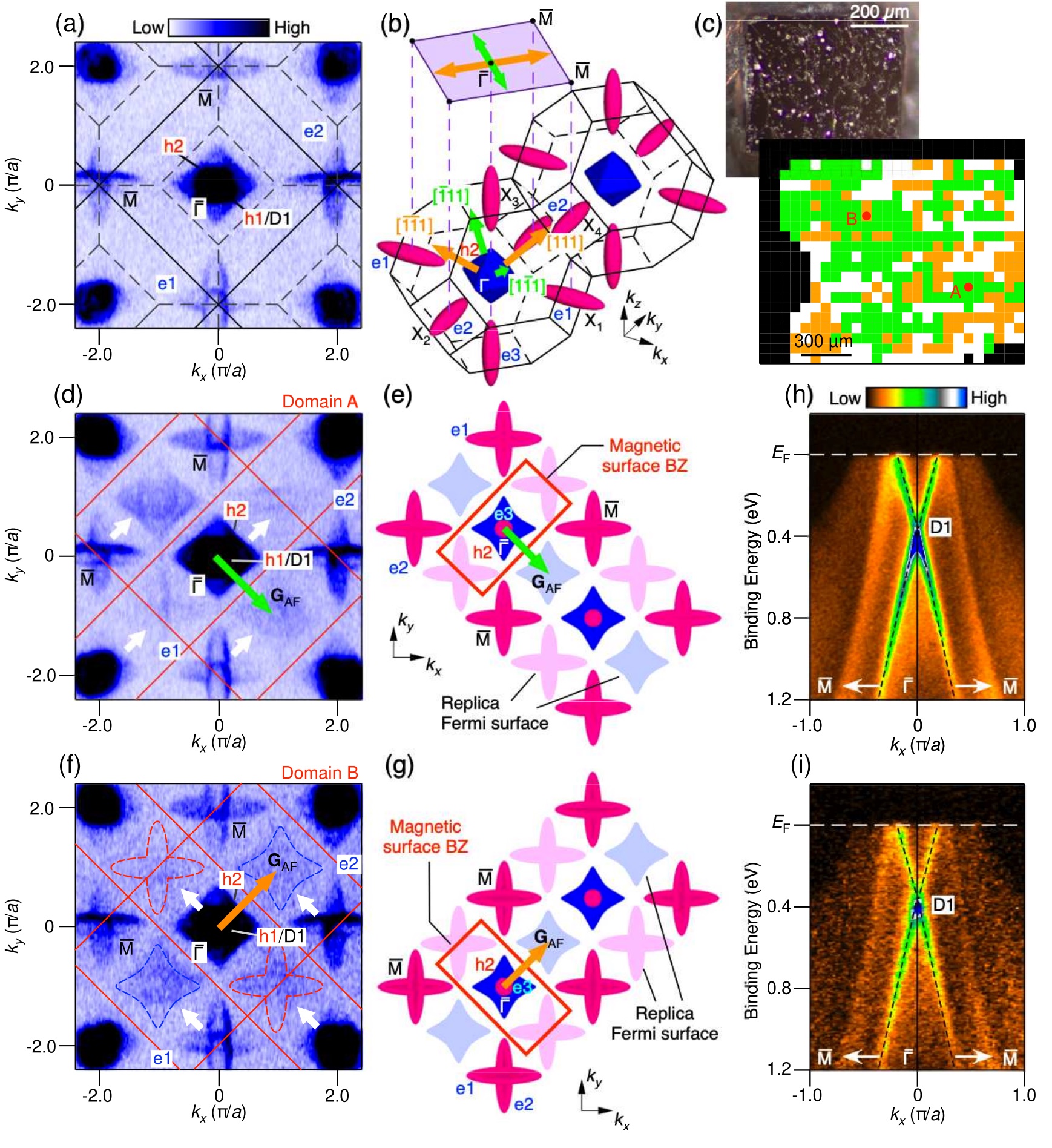}
\caption{\label{FIG2}(color online). (a) ARPES-intensity mapping at $E_\mathrm{F}$ for GdBi plotted as a function of $k_x$ and $k_y$ at $T = 40\ \mathrm{K}$ obtained with $h\nu = 105\ \mathrm{eV}$. Solid lines indicate the surface BZ boundary whereas the dashed lines indicate the cross-section of both $k_z = 0$ and $\pi$ planes of the bulk fcc BZ boundary. (b) Bulk fcc BZ together with the bulk Fermi surface. Purple rectangle indicates the surface BZ projected onto the (001) surface. Orange and green arrows indicate the type-II AF ordering vectors in bulk BZ and projected surface BZ, respectively. (c) (Top) microscope image of the cleaved surface of GdBi, and (bottom) distribution of AF domains visualized by scanning ARPES with $26\times22$ mesh. The pixel size is $50\times50\ \mathrm{\mu m^2}$. Regions shown with orange and green color correspond to domains A and B where the AF ordering vector points to lower right and upper right directions, respectively. White region represents the area where the domain identification was difficult, while the black region has no signal from the sample. (d) Same as a but obtained at $T = 6\ \mathrm{K}$ (AF phase) for domain A. $\mathrm{\mathbf{G}_{AF}}$ (green arrow) is the AF ordering vector projected onto the surface. White arrows indicate the replica of Fermi surfaces and red rectangles represent the magnetic surface BZ. (e) Schematics of band folding in the surface BZ, together with the folded Fermi surfaces for domain A. Red rectangle represents the magnetic surface BZ. (f), (g) Same as (d) and (e), respectively, but for domain B. (h), (i) Near-$E_\mathrm{F}$ ARPES-intensity plots around the $\bar{\mathrm{\Gamma}}$ point obtained at $T = 6\ \mathrm{K}$ along the $k_x$ cut for domains A and B, respectively.
}
\end{center}
\end{figure}

\begin{figure}
\begin{center}
\includegraphics[width=6in]{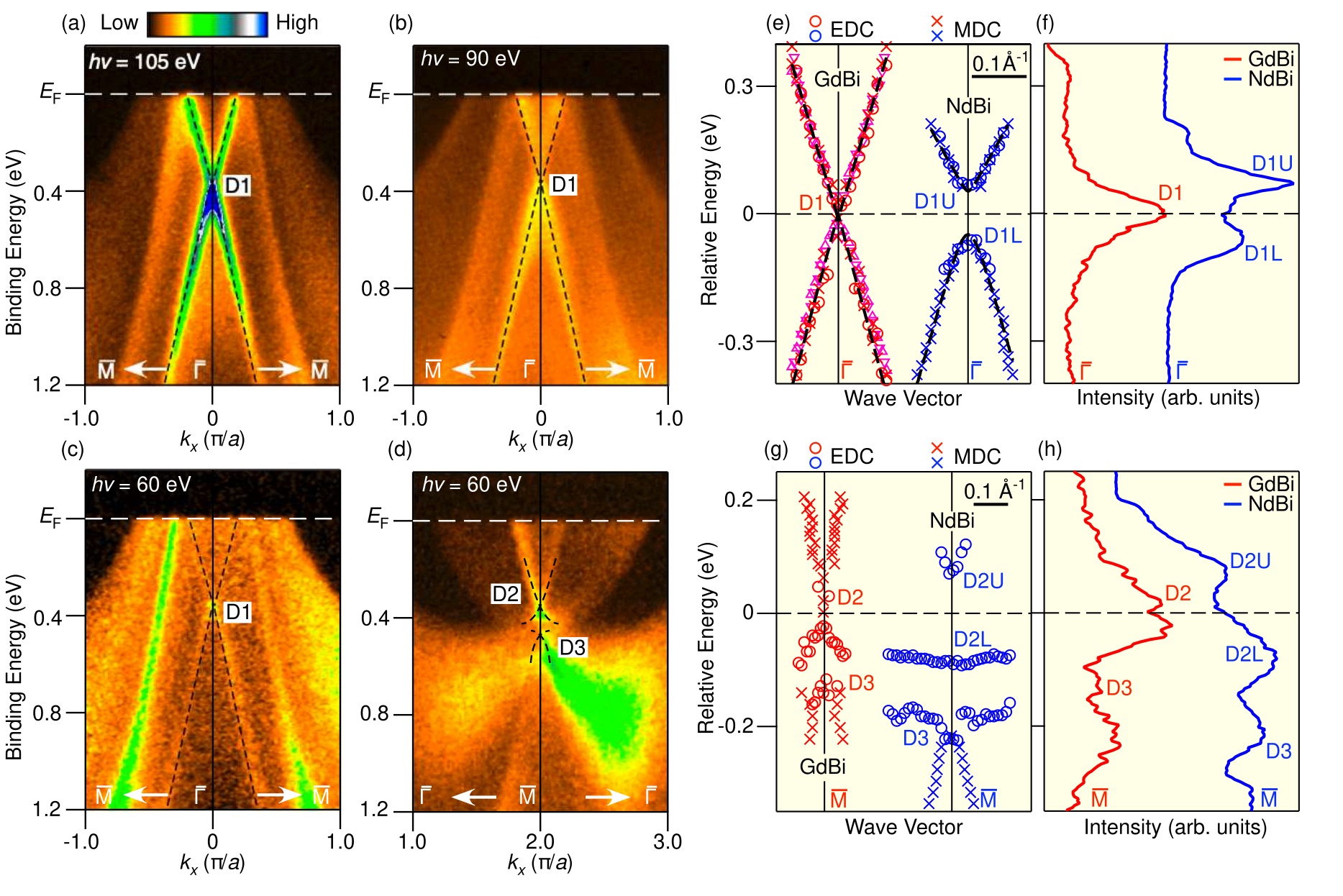}
  \hspace{0.2in}
\caption{\label{FIG3}(color online). (a)–(c) Plots of ARPES intensity around the $\bar{\mathrm{\Gamma}}$ point for GdBi along the $k_x$ axis obtained at $T = 7\ \mathrm{K}$ with $h\nu = 105, 90, \mathrm{and}\ 60\ \mathrm{eV}$, respectively. (d) Plot of ARPES intensity around the $\bar{\mathrm{M}}$ point for GdBi obtained at $T = 7\ \mathrm{K}$ with $h\nu = 60\ \mathrm{eV}$. (e) Experimental D1-band dispersion for GdBi and NdBi in the AF phase. Open circles and cross markers represent the band dispersion obtained by tracing the peak position of energy distribution curves (EDCs) and momentum distribution curves (MDCs), respectively, measured with $h\nu = 90\ \mathrm{eV}$. Magenta triangles represent the EDC/MDC peak plot for the data obtained with $h\nu = 105\ \mathrm{eV}$. D1U and D1L stand for upper and lower branches of gapped D1 band, respectively. (f) EDCs for GdBi (red curve) and NdBi (blue curve) at the $\bar{\mathrm{\Gamma}}$ point in which the energy zero is set to be at the Dirac point of the D1 band in the PM phase. (g), (h) Same as (e) and (f), respectively, but for the D2 and D3 bands in which the energy zero is aligned to the Dirac point of the D2 band in the PM phase. }
\end{center}
\end{figure}

\begin{figure}
\begin{center}
\includegraphics[width=6in]{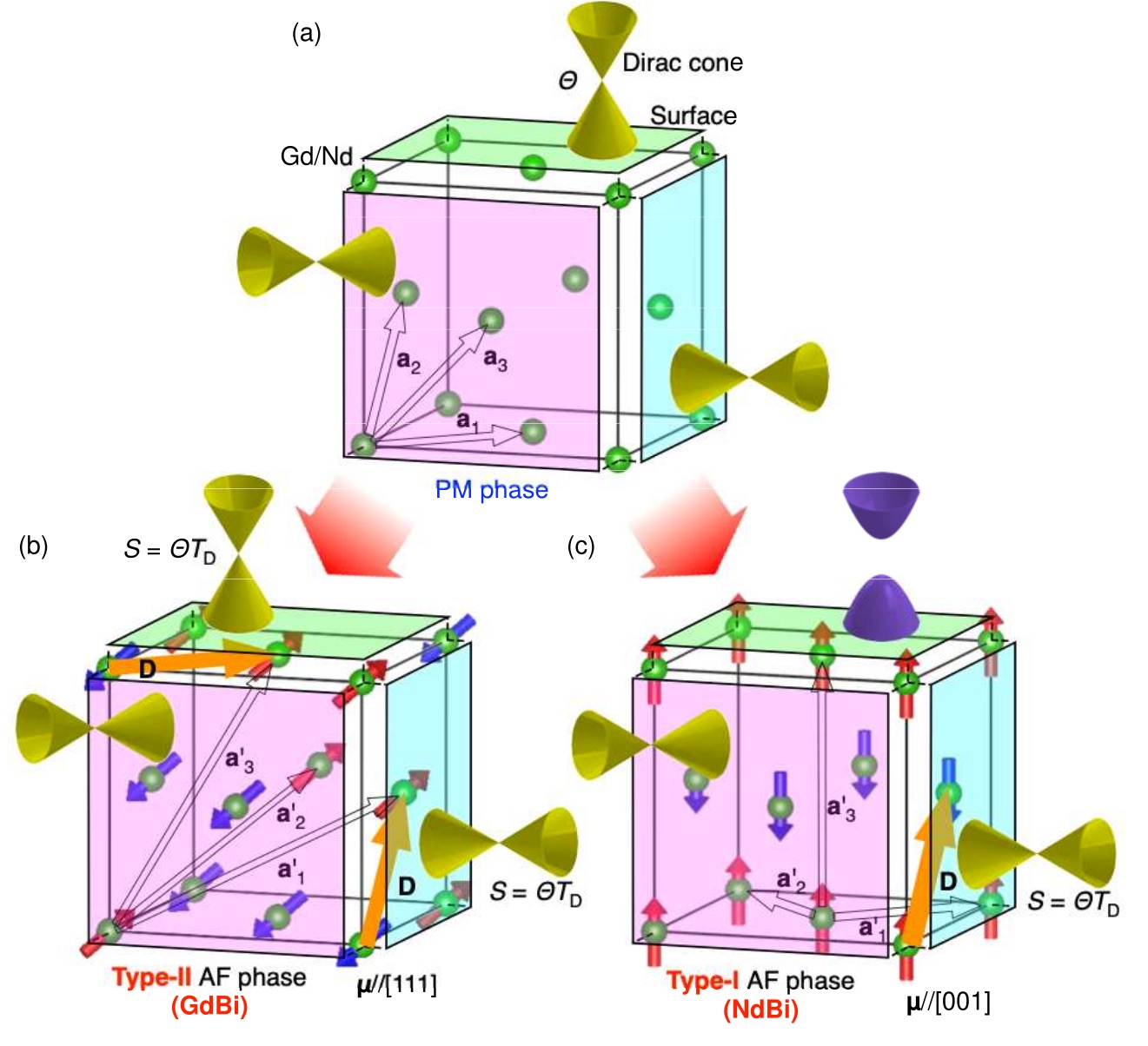}
  \hspace{0.2in}
\caption{\label{FIG4}(color online). (a) Schematic of surface Dirac cones in the PM phase for both NdBi and GdBi, which are attributed to the strong TI phase with surface-independent massless Dirac cones protected by TRS. (b), (c) Same as a but for the type-II (GdBi) and type-I (NdBi) AF phases, respectively. Dirac cones are protected by the combined symmetry \textit{S}, except for the ferromagnetically aligned surface for the type-I case. The unit lattice vectors $\mathrm{\mathbf{a}}’_i$ ($i = 1–3$) that take into account the magnetic structure for the AF case are shown by open black arrows. \textbf{D} vectors are shown by orange color in (b) and (c). 
}
\end{center}
\end{figure}

\begin{figure}
\begin{center}
\includegraphics[width=6in]{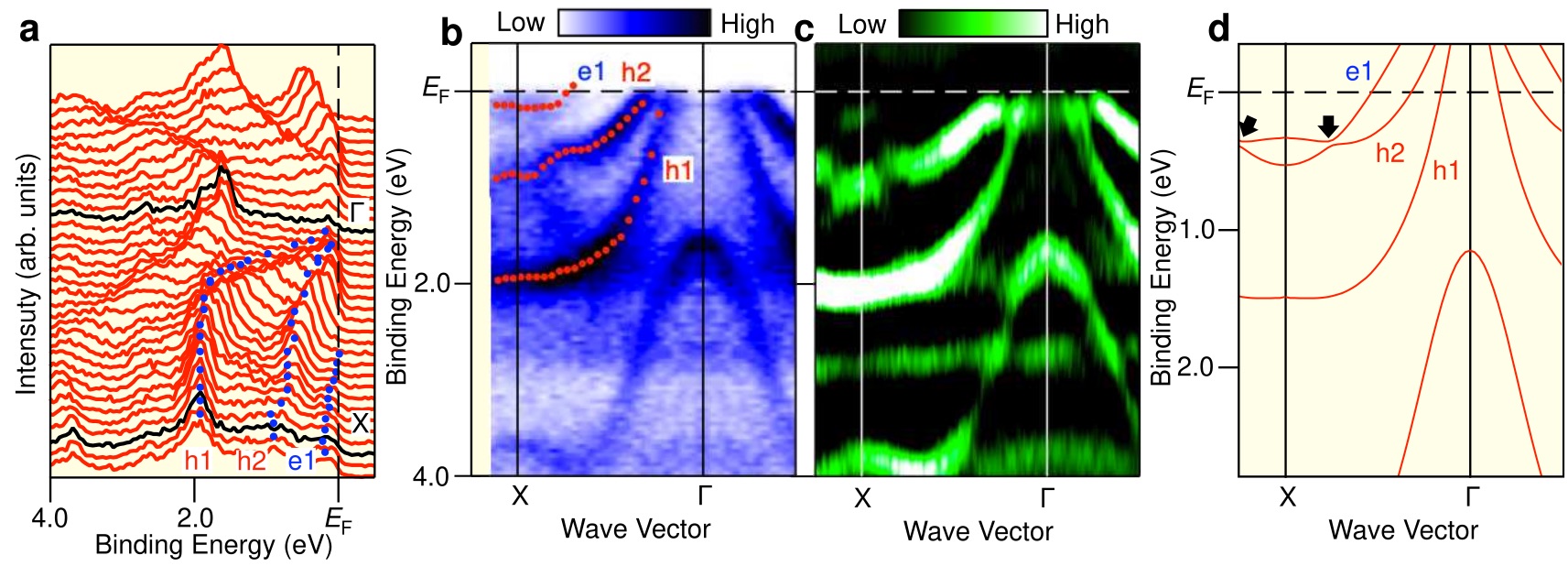}
  \hspace{0.2in}
\caption{\label{FIG5}(color online). (a) Energy distribution curves of GdBi measured along the $\mathrm{\Gamma X}$ cut of bulk BZ at $T = 40\ \mathrm{K}$ with $h\nu = 536\ \mathrm{eV}$. (b), (c) Corresponding raw and second derivative ARPES intensities, respectively, plotted as a function of $k_x$ and binding energy. Red and blue dots in (a) and (b) are a guide for the eyes to trace the Bi-6$p$ (h1 and h2) and Gd 5$d$ (e1) bands, respectively. (d) Calculated band structure along the $\mathrm{\Gamma X}$ cut of bulk BZ, vertically expanded by a factor of 1.45 to obtain a better matching with the experiment.
}
\end{center}
\end{figure}

\begin{figure}
\begin{center}
\includegraphics[width=4.2in]{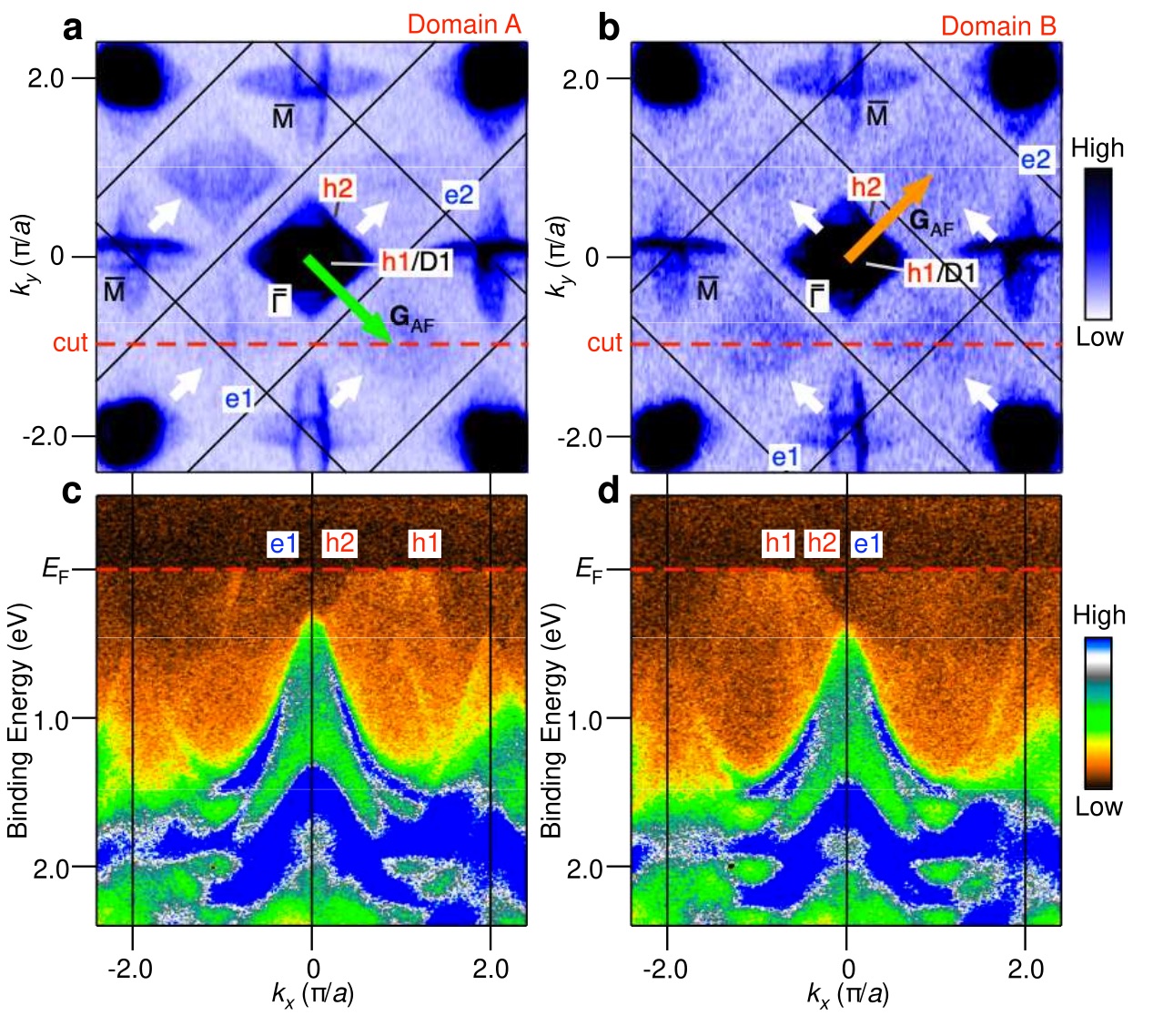}
\hspace{0.2in}
\caption{\label{FIG6}(color online). (a),(b) Fermi surface mapping at $E_\mathrm{F}$ of GdBi at $T = 6\ \mathrm{K}$ for domains A and B, respectively [same as Figs. 2(d) and 2(f)], but plotted without guidelines of replica Fermi surfaces. (c), (d) ARPES intensities measured along a $k$ cut shown by a red dashed line in (a) and (b), respectively.
}
\end{center}
\end{figure}

{There exist two types of magnetic structures (type A and type B) in the type-II antiferromagnet \cite{YYLi1955PRB}. Type A has a magnetic structure shown in Fig.\ \hyperref[FIG1]{1(a)} in which adjacent ferromagnetic layers in the (111) planes align in an antiparallel way, producing the 2 $\times$ 1 Fermi-surface reconstruction on the (001) cleaved surface. On the other hand, in type B, magnetic moment in the (111) plane does not align ferromagnetically, but next nearest neighbor atoms couple with antiparallel configuration as in the case of type A; in this case, 2 $\times$ 2 reconstruction of the Fermi surface takes place. While previous studies did not clearly distinguish type A or type B, the present ARPES measurement clearly identifies the 2 $\times$ 1 Fermi-surface reconstruction, strongly supporting the type-A configuration.}

\bibliography{GdBi_r.bib}

\clearpage

\clearpage

\clearpage

\clearpage

\clearpage

\clearpage

\end{document}